\documentclass[twocolumn,showpacs,prb,superscriptaddress,amsmath,amssymb]{revtex4}
\usepackage{graphicx}     
\begin{document}

\title{Theory of Coherent Tunneling in Exciton Condensates
of Bilayer Quantum Hall Systems}
\author{K. Park}
\affiliation{School of Physics, Korea Institute for Advanced
Study, Seoul 130-722, Korea}
\author{S. Das Sarma}
\affiliation{Condensed Matter Theory Center, Department of
Physics, University of Maryland, College Park, MD 20742-4111}
\date{\today}

\begin{abstract}
Due to strong interlayer correlations, the bilayer quantum Hall
system is a single coherent system as a whole rather than a
weakly-coupled set of two independent systems, which makes
conventional tunneling theories inapplicable. In this paper, we
develop a theory of interlayer tunneling in coherent exciton
condensates of bilayer quantum Hall systems at total filling
factor $\nu_T=1$. One of the most important consequences of our
theory is that the zero-bias interlayer tunneling conductance peak
is strongly enhanced, but fundamentally finite even at zero
temperature. We explicitly compute the height of the conductance
peak as a function of interlayer distance, which is compared with
experiment. It is emphasized that the interlayer distance
dependence of the conductance peak is one of the key properties
distinguishing between the spontaneous coherence due to many-body
effects of the Coulomb interaction and the induced coherence due
to the single-particle tunneling gap. It is also emphasized that,
though the strongly enhanced tunneling conductance originates from
the interlayer phase coherence, it is not the usual Josephson
effect. We propose an experimental setup for the true Josephson
effect in couterflowing current measurements for a coupled set of
two bilayer quantum Hall systems, which is a more precise analogy
with the real Josephson effect in superconductivity.
\end{abstract}

\maketitle

\section{Introduction}
\label{introduction}

Strongly-enhanced interlayer conductance peaks observed by
Spielman {\it et al.} \cite{Spielman} near zero bias in bilayer
quantum Hall systems at total filling factor $\nu_T=1$ have
attracted much interest for many reasons. One of the main reasons
is the existence of a rather precise analogy between the bilayer
quantum Hall effect \cite{Murphy} and superconductivity. That is,
the ground state of the bilayer quantum Hall effect at small
interlayer distance $d/l_B \ll 1$ ($l_B=\sqrt{\hbar c/eB}$) can be
mapped onto the BCS-type wavefunction describing the Bose-Einstein
condensate of exciton pairs formed between electrons and holes
residing across the interlayer barrier.

Bose-Einstein condensation of neutral excitons has, in fact, been
sought after in semiconductors for decades. In particular, there
have been fascinating recent experiments on possible condensation
of optically generated indirect excitons \cite{Indirect}, for
which, however, evidence is not yet conclusive. On the other hand,
in the bilayer quantum Hall effect, it is generally accepted that
the strongly-enhanced conductance peak is a direct indication of
the macroscopic phase coherence.

To be concrete with respect to the mapping between
superconductivity and the bilayer quantum Hall effect, let us
begin by writing the ground state wave function at $d/l_B
\rightarrow 0$. The explicit ground state wave function at $d/l_B
\rightarrow 0$ is very instructive in illustrating the essential
physics even though the ground state is substantially more
complicated for general $d/l_B$, especially near the critical
$d/l_B$ where the exciton condensate disappears. (Note that we
will use numerical methods to study the ground state at general
$d/l_B$.) The ground state wave function at $d/l_B \rightarrow 0$
is known as the Halperin's (1,1,1) state \cite{Halperin111}:
\begin{equation}
|\psi_{111}\rangle = \prod_{m}
(c^{\dagger}_{m\uparrow} + c^{\dagger}_{m\downarrow}) |0\rangle ,
\label{111}
\end{equation}
where $m$ is a momentum index of the lowest Landau level. In the
above, the pseudospin representation is used: $\uparrow$ and
$\downarrow$ indicate the top and the bottom layer, respectively.
It is important to note that, contrary to the usual representation
of the (1,1,1) state,
\begin{eqnarray}
\psi_{111}=\prod_{i,j\in\uparrow}(z_i -
z_j)\prod_{k,l\in\downarrow}(z_k - z_l) \prod_{m\in\uparrow
,n\in\downarrow}(z_m - z_n)
\end{eqnarray}
which denotes only orbital components, Eq.(\ref{111}) describes
the full wave function including both the orbital and the layer
degree of freedom \cite{Yang}.

The ground state wave function described by Eq.(\ref{111}) has an
isomorphic structure to the BCS wave function, which can be made
more apparent after the following reorganization:
\begin{eqnarray}
|\psi_{111}\rangle &=& \prod_{m} (1+ c^{\dagger}_{m\uparrow}
c_{m\downarrow}) \prod_{m'} c^{\dagger}_{m'\downarrow} |0\rangle
\nonumber \\
&=& \prod_{m} (1+ c^{\dagger}_{m\uparrow} c_{m\downarrow})
|\textrm{new vacuum}\rangle
\nonumber \\
&=& \prod_{m} (1+ c^{\dagger}_{m\uparrow}
h^{\dagger}_{m\downarrow}) |\textrm{new vacuum}\rangle,
\end{eqnarray}
where $h^{\dagger}$ is a creation operator for holes, acting on
the fully-filled bottom layer. Because of this analogy, it is
rather natural to expect that the bilayer quantum Hall state may
have a coherence in the phase associated with interlayer
electron-number difference, which is similar to the phase
coherence associated with the total number of Cooper pairs in
superconductivity. The phase coherence between states with
different Cooper-pair numbers is the origin of the Josephson
effect in superconductivity. Naturally, this similarity led many
previous authors \cite{WenZee,Ezawa} to predict the Josephson
effect in bilayer quantum Hall systems. So, in this context, the
strongly enhanced conductance observed by Spielman {\it et al.}
\cite{Spielman} seemed to be exactly the experimental verification
needed. There are, however, some important properties of the
conductance peak indicating that this phenomenon is not the
conventional Josephson effect; most notably, experiments suggest
that the height and the width of the zero-bias conductance peaks
saturate to finite values when available finite-temperature data
are extrapolated to the zero temperature limit
\cite{Spielman2,Eisenstein}. This means that there is no DC
current strictly at zero bias voltage in contrast to the real
Josephson effect in superconductivity.

This apparent discrepancy gave rise to two groups of thought. In
one group, the enhanced conductance is still regarded as DC
Josephson effect, but its height is reduced by complicated
disorder-induced fluctuations \cite{Stern1,Stern2,Balents,Fogler}.
On the other hand, others \cite{Joglekar} have argued that there
is no exact analog of the Josephson effect in the experimental
setup measuring interlayer tunneling currents in a single set of
bilayer quantum Hall systems. It is because the bilayer system as
a whole is a single Bose-Einstein condensate(BEC), not a set of
two independent BEC's. While this argument itself is clearly true,
it is still necessary to explain why and how the enhanced
interlayer conductance is related to the BEC of excitons. The
resolution to this crucial issue has remained elusive even after a
very extensive body of research efforts continuing to recent years
\cite{FS,Wang,DKKLee,KD,FM,RNM}.

Important issues to be addressed are, in particular, (i) exactly
what is the process of coherent tunneling in interlayer tunneling
current measurements, (ii) how this process can be mathematically
formulated, and finally (iii) what is the precise relationship
between the interlayer phase coherence and the zero-bias
conductance. It is the goal of this paper \cite{preprintPark} to
address these questions. Specifically, we would like to provide a
quantitative analysis of the zero-bias conductance peak as a
function of $d/l_B$, which, in turn, generates a sharp distinction
between the coherent and incoherent tunneling processes.

The $d/l_B$ dependence of the conductance peak is important also
because it distinguishes between the coherence due to many-body
effects of the Coulomb interaction and that due to the
single-particle tunneling. In this paper, we are primarily
interested in the regime of sufficiently small single-particle
tunneling, $t/(e^2/\epsilon l_B) \ll 1$. We are interested in this
regime mainly because experimental values of $t/(e^2/\epsilon
l_B)$ are really quite small; in recent experimental setups, it
can be as small as $10^{-6} - 10^{-7}$. This limit is also quite
interesting in a theoretical point of view since the interlayer
phase coherence is induced purely by many-body effects of the
Coulomb interaction in the absence of single-particle tunneling.
In later sections, we will investigate similarities and
differences between the spontaneous coherence caused by the
Coulomb interaction and the induced coherence caused by
single-particle tunneling.

This paper is organized as follows; we begin in Section
\ref{coherent} by distinguishing physical situations for coherent
tunneling from those for incoherent one. We provide a physical
picture as to exactly how interlayer tunneling can be enhanced in
the presence of coherence. We, then, present our quantitative
analysis in Section \ref{Hamiltonian} by providing a mathematical
derivation of the many-body tunneling Hamiltonian and its
connection to the interlayer tunneling conductance. In Section
\ref{numerics}, we numerically evaluate the coherent tunneling
conductance as a function of $d/l_B$ and compare it with
experiment. In Section \ref{realJosephson}, we propose an
experimental setup for the Josepson effect in counterflowing
current measurements for a coupled set of two bilayer quantum Hall
systems. We finally conclude in Sec. \ref{conclusion}.

\section{Coherent tunneling versus incoherent tunneling}
\label{coherent}

\begin{figure}
\includegraphics[width=3.5in]{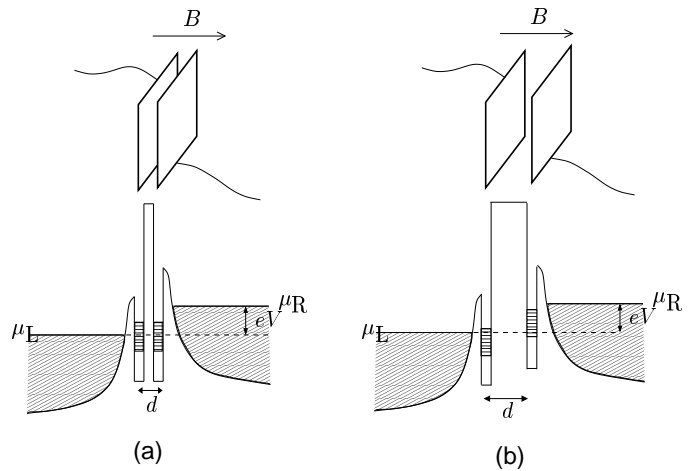}
\caption{Diagram showing the difference between (a) coherent
tunneling and (b) incoherent tunneling. Note that, for the
coherent ground state, there is no chemical potential difference
between the two layers, as depicted in (a). When $d$ is
sufficiently large as shown in (b), however, each layer becomes
independent, and therefore the interlayer coherence is lost.
\label{coherent_incoherent}}
\end{figure}
As mentioned in the introduction, the bilayer quantum Hall system
itself is a single superfluid system. The very existence of the
excitation gap required for the quantized Hall resistance
indicates that the ground state wave function is robust against
small perturbations, in which case the bilayer system can be
assumed to maintain its equilibrium. External perturbations such
as bias voltage, then, can be taken as perturbations to the
bilayer system as a whole, not to the individual layers. In this
situation, there is no chemical potential difference between the
two layers, as depicted in Fig.\ref{coherent_incoherent} (a).

\begin{figure}
\includegraphics[width=2.2in,angle=0]{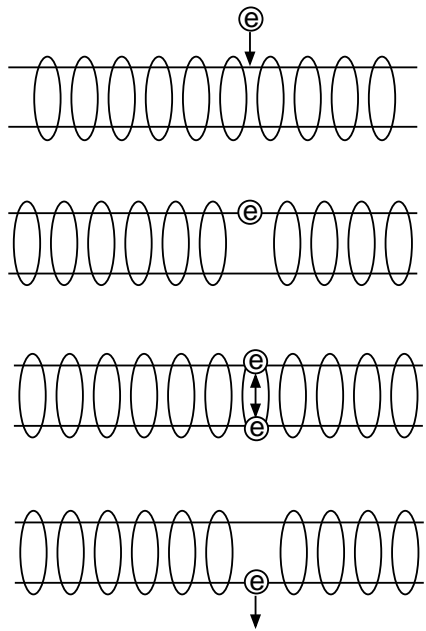}
\caption{Schematic diagrams for coherent interlayer tunneling.
Note that each ellipse indicates the resonance between the top and
the bottom layer. Imagine that an electron is inserted into the
top layer of the bilayer quantum Hall system, as depicted in the
top diagram. The inserted electron, then, becomes a quasiparticle
on top of the ground state with an energy penalty (the second
diagram from the top). The so-introduced quasiparticle, then,
immediately becomes a part of the coherent ground state by
resonating between the two layers (the third diagram from the
top). It is important to note that the interlayer resonance can be
caused either by many-body effects of the Coulomb interaction or
by the single-particle tunneling. Finally, the electron is ejected
from the bottom layer, and thereby the interlayer current flows
(the bottom diagram). \label{schematic}}
\end{figure}

In the situation described by Fig.\ref{coherent_incoherent} (a),
tunneling is coherent because the aforementioned robustness of the
ground state is due to the interlayer coherence. Physically
speaking, coherent interlayer tunneling is obtained as follows:
Imagine that an electron is inserted into the top layer of the
bilayer quantum Hall system, as depicted in the top diagram in
Fig.\ref{schematic}. The inserted electron, then, becomes a
quasiparticle on top of the ground state with an energy penalty
(the second diagram from the top in Fig.\ref{schematic}). The
energy penalty can be caused either by the Coulomb interaction
(for the spontaneous coherence) or by the single-particle
tunneling gap (for the induced coherence). The so-introduced
quasiparticle, then, immediately becomes a part of the coherent
ground state by resonating between the two layers (the third
diagram from the top). The electron is finally ejected from the
bottom layer, and thereby the interlayer current flows. It is
important to note that, when there is an interlayer coherence,
tunneling can be strongly enhanced because all inserted electrons
will tunnel in a synchronized fashion.


The situation is quite different when there is no interlayer
coherence. There are two important cases where the interlayer
coherence can be lost; (i) for a fixed filling factor, namely
$\nu_T=1$ in our case, $d/l_B$ becomes large or (ii) simply $B=0$
in which interactions play a weaker role. For the first case, each
layer forms an individual composite Fermi sea
\cite{Jain,Fradkin,Halperin} at filling factor $\nu=1/2$. What
actually tunnel between the layers, however, are real electrons
rather than composite fermions. In this situation, the interlayer
tunneling is suppressed due to the fact that the composite Fermi
sea is a highly correlated state so that the sudden insertion of
an uncorrelated electron requires a large energy cost.

When $B=0$, the situation is simpler: ordinary electrons tunnel.
In this situation, the interlayer tunneling current can be
evaluated in a conventional perturbation theory assuming that the
single-particle tunneling gap, $t$, is small and therefore the two
layers become independent and have individual chemical potentials,
as depicted in Fig.\ref{schematic} (b). To be concrete, let us
start with the tunneling Hamiltonian, $H_t$:
\begin{eqnarray}
H_t &=& t \sum_{\bf k} ( c^{\dagger}_{{\bf k}\uparrow} c_{{\bf k}
\downarrow} + c^{\dagger}_{{\bf k}\downarrow}c_{{\bf k}\uparrow})
\end{eqnarray}
where $t$ is the single-particle tunneling gap and the pseudospin
representation is used. It is assumed that momenta parallel to the
two-dimensional plane are conserved; in other words, there is no
tunneling between different ${\bf k}$'s. In this representation,
it is not too difficult to show that the tunneling current
operator, $\hat{J}$, is given as follows:
\begin{eqnarray}
\hat{J} &=& et i \sum_{\bf k} ( c^{\dagger}_{{\bf k}\uparrow}
c_{{\bf k}\downarrow} - c^{\dagger}_{{\bf k} \downarrow}c_{{\bf k}
\uparrow})
\end{eqnarray}
When $t$ is small, one can use a conventional first-order S-matrix
expansion to compute the expectation value of the current
operator:
\begin{equation}
I(t)= -i \int^{t}_{-\infty} d t' \langle [\hat{J}(t), H_T(t')]
\rangle ,
\end{equation}
which, after some algebra, leads to the following expression:
\begin{eqnarray}
I = 2e t^2 \int \frac{d\varepsilon}{2\pi} \sum_{\bf k} \rho_{{\bf
k} \uparrow}(\varepsilon) \rho_{{\bf k}\downarrow}(\varepsilon+eV)
[f(\varepsilon)-f(\varepsilon+eV)] \label{normal}
\end{eqnarray}
where $\rho_{\bf k}(\varepsilon)$ is the density of states for a
given {\bf k} (which is also known the spectral function), and
$f(\varepsilon)$ is the usual Fermi-Dirac distribution function.
Note that $\rho_{{\bf k} \uparrow}(\varepsilon) = \rho_{{\bf
k}\downarrow}(\varepsilon)\equiv\rho_{\bf k}(\varepsilon)$ because
of symmetry. Now, since the $z$-component motion of electrons is
frozen due to the confinement to the two-dimensional plane,
$\rho_{\bf k} (\varepsilon)$ is sharply peaked around its center
(which depends on ${\bf k}$). As a consequence, the integrand in
Eq.(\ref{normal}) vanishes except for small $eV$:
\begin{eqnarray}
\rho_{\bf k}(\varepsilon)\rho_{\bf k}(\varepsilon+eV) \simeq
\rho^2_{\bf k}(\varepsilon) e^{-{(eV)}^2/2\delta^2},
\end{eqnarray}
where the size of $\delta$ is determined by the static disorder,
which controls the width of the quasiparticle coherence peak. For
small $\delta$, as is expected in high-mobility 2D systems of
interest, this single-particle tunneling peak is almost a
delta-function with a small resonance width $eV \simeq \delta$.

This resonance behavior of the interlayer tunneling conductance
is, in fact, observed at $B=0$ in the same sample \cite{Spielman}
showing the enhancement of the zero-bias tunneling conductance at
$\nu_T=1$ for small $d/l_B$. So, at least superficially,
incoherent tunneling seems to generate similar physical
consequences as coherent tunneling. There are, however,
indications suggesting that the $B=0$ conductance peak is actually
different from the $\nu_T=1$ peak. One indication is that the
$B=0$ conductance peak is 100 times smaller than the peak at
$\nu_T=1$. Furthermore, the $B=0$ peak is at zero bias only when
the two layers have the same density. If the densities are
different, the $B=0$ peak shifts accordingly. In contrast, the
$\nu_T=1$ peak is locked on to $V=0$, even for (small) non-zero
density differences. Despite these indications, however, it is
very important to know whether coherent tunneling can give rise to
results which are manifestly physically distinct from the
corresponding incoherent tunneling results, and, if so, what those
results are.

One of such distinctions can be obtained in the $d/l_B$ dependence
of the zero-bias tunneling conductance. In Eq.(\ref{normal}), the
only dependence of the incoherent tunneling current on $d$ is
through the tunneling gap, $t$. Experimentally, $d/l_B$ is varied
usually by changing the magnetic field for the same physical
sample (Note that the electron density is also adjusted to
maintain the fixed filling factor). In this experimental setup,
$t$ should not change much since $t$ depends on the physical
distance $d$, not on $d/l_B$. Experimentally observed, however, is
an abrupt change in the zero-bias tunneling conductance: the
zero-bias conductance peak completely disappears after a certain
critical distance $d_c/l_B$ \cite{Spielman}. This is the reason
why it is so important to develop a quantitative theory to compute
the zero-bias conductance peak as a function of $d/l_B$. We
provide such a theory in the next section. We believe that the key
to understanding the interlayer coherent tunneling experiment of
Spielman {\it et al.}\cite{Spielman} is its $d/l_B$ dependence,
which, as a matter of principle, can distinguish between
incoherent single-particle tunneling and the many-body correlation
induced coherent effect of bilayer excitonic condensates.


\section{Derivation of the many-body tunneling Hamiltonian}
\label{Hamiltonian}

In this section, we develop a theory of coherent tunneling which
occurs when there is an interlayer coherence in the ground state.
An important starting point is that the two layers cannot be
regarded as independent systems. Instead, they must be treated as
a coherent whole. Since there is no interlayer chemical potential
in a single coherent bilayer system, there is no electromotive
force within the bilayer system and therefore any current should
be induced from outside. So, it is necessary to take into account
external leads, as schematically shown in
Fig.\ref{coherent_incoherent}. This, of course, makes any
quantitative prediction dependent on the way in which bilayer
systems are connected to external leads. However, it is still
possible to make quantitative predictions on some essential
aspects of coherent interlayer tunneling. In particular, we study
the $d/l_B$ dependence of the zero-bias tunneling conductance
peak. In doing so, we also show that the width of the conductance
peak is fundamentally finite even at zero temperature, and it is
controlled ultimately by very small, but finite single-particle
interlayer tunneling gap $t$. This makes sense since a finite $t$
is essential to experimentally drive an interlayer current.

Let us begin our quantitative analysis by writing the total
Hamiltonian including (i) the interlayer (single-particle)
tunneling Hamiltonian, $H_t$, (ii) the Hamiltonian for the Coulomb
interaction between electrons, $H_\textrm{Coul}$, (iii) the
Hamiltonian describing the left and the right lead, $H_L$ and
$H_R$ respectively, and finally (iv) the Hamiltonian for tunneling
between leads and the bilayer system, $H'$:
\begin{eqnarray}
H &=& H_0 + H' +H_R +H_L,
\\
H_0 &=& H_t +H_\textrm{Coul},
\\
H_t &=& t \sum_m ( c^{\dagger}_{m\uparrow} c_{m\downarrow} +
c^{\dagger}_{m\downarrow}c_{m\uparrow}) \equiv 2 t S_x,
\label{H_t}
\\
\frac{H_\textrm{Coul}}{e^2/\epsilon l_B} &=&
{\cal P}_{LLL} \Big( \sum_{i,j\in\uparrow} \frac{1}{r_{ij}}
+\sum_{k,l\in\downarrow} \frac{1}{r_{kl}}
\nonumber \\
&+&\sum_{i\in\uparrow,k\in\downarrow}
\frac{1}{\sqrt{r^2_{ik}+(d/l_B)^2}}
\Big) {\cal P}_{LLL},
\\
H' &=& \sum_{k,m} T_{R\uparrow}(k,m)
[c^{\dagger}_R(k)c_{m\uparrow} + \textrm{H.c.}]
\nonumber \\
&+& \sum_{p,m'} T_{L\downarrow}(p,m')
[c^{\dagger}_L(p)c_{m'\downarrow} + \textrm{H.c.}], \label{H'}
\end{eqnarray}
where, as before, the pseudospin representation is used and $m$ is
a momentum index of the lowest Landau level. ${\cal P}_{LLL}$ is
the lowest Landau level projection operator. $T_{R\uparrow}(k,m)$
is the tunneling amplitude between the state with momentum $k$ in
the right lead, and the state with $m$ in the top layer of the
bilayer system. $T_{L\downarrow}(p,m')$ is similarly defined.
$H_R$ and $H_L$ describe external leads as normal Fermi liquids.
Note that $H_t$ is related to the pseudospin magneization in the
$x$-direction, $S_x$:
\begin{eqnarray}
S_x=\frac{1}{2} \sum_m ( c^{\dagger}_{m\uparrow} c_{m\downarrow} +
c^{\dagger}_{m\downarrow}c_{m\uparrow}).
\end{eqnarray}
Also, note that the internal interlayer current operator is
related to the pseudospin magneization in the $y$-direction,
$S_y$:
\begin{eqnarray}
\hat{J}_\textrm{inter}= eti \sum_m ( c^{\dagger}_{m\uparrow}
c_{m\downarrow} - c^{\dagger}_{m\downarrow}c_{m\uparrow}) \equiv
2etS_y.
\end{eqnarray}

Now that we are interested in the limit of zero $t/(e^2/\epsilon
l_B)$, we investigate the possibility of a current operator which
survives as $t/(e^2/\epsilon l_B) \rightarrow 0$. This operator
will, then, measure the spontaneous interlayer current as opposed
to the single-particle hopping induced current. To this end, let
us consider the following tunneling Hamiltonian:
\begin{equation}
H'_T = H' \frac{1}{E_g-H_0-H_R-H_L} H', \label{H'_T}
\end{equation}
where $E_g$ is the ground state energy of $H_0+H_R+H_L$. The idea
is that, by adding an electron to the top layer and removing
another from the bottom layer, $H'_T$ describes the total current
flowing through the bilayer system. Note that $H'_T$ is second
order in $H'$ since it is the lowest order that can carry the
current from one lead to the other lead in the limit of zero
$t/(e^2/\epsilon l_B)$.

Normally, the above process would not cause actual interlayer
tunneling, but instead it would result in charge build-up. Our
bilayer quantum Hall system, however, is special such that it has
spontaneous interlayer coherence for small interlayer distances
and therefore electrons can move back and forth between the layers
even in the limit of zero $t/(e^2/\epsilon l_B)$. The total
interlayer current is, of course, zero in the ground state since
the back and forth currents cancel each other. Non-zero interlayer
currents, however, can be induced by applying a bias voltage which
breaks the balance between the back and forth motion in
equilibrium.

Once interlayer currents begin to flow as described by the process
of $H'_T$ in Eq.(\ref{H'_T}), the currents can flow steady only
when it is carried by the internal interlayer current: $I=\langle
\hat{J}_\textrm{inter} \rangle$, as dictated by the charge
conservation. The exciton condensate can carry the interlayer
current by adjusting its interlayer phase difference $\phi$. To
understand this, it is instructive to consider the ground state
wavefunction at $d/l_B \rightarrow 0$ in the presence of non-zero
interlayer phase:
\begin{equation}
|\psi_{111}(\phi)\rangle = \prod_{m} (c^{\dagger}_{m\uparrow} +
e^{i\phi} c^{\dagger}_{m\downarrow}) |0\rangle . \label{111phi}
\end{equation}
It can, then, be shown that the ground state in Eq.(\ref{111phi})
carries a non-zero interlayer current:
\begin{equation}
\langle \hat{J}_\textrm{inter} \rangle =2et \langle S_y \rangle =
2et \langle S \rangle \sin{\phi}, \label{Jinter}
\end{equation}
where $\langle S \rangle \equiv |\langle \sum_m
c^{\dagger}_{m\uparrow}c_{m\downarrow} \rangle |$.

It is important that there is a maximum current that can be
carried by the coherent state:
\begin{equation}
\langle \hat{J}_\textrm{inter} \rangle \leq J_\textrm{critical} =
2et \langle S \rangle. \label{Jcritical}
\end{equation}
The existence of the maximum coherent current is analogous to that
of the critical current in the superconductivity. This termination
of the coherent tunneling current beyond a critical value is the
reason why the tunneling conductance has a narrow peak near zero
bias. That is, the tunneling current is coherent only within a
small window of bias voltage. Once the bias voltage gets larger
than a critical value, the tunneling current becomes too large to
be carried only by coherent tunneling. The excess current must,
then, be carried by incoherent tunneling, which has a low
conductance, as argued previously in terms of the formation of the
composite fermion sea.

It is important to note that $H'_T$ in Eq.(16) is zeroth order in
$t$ whereas, in usual incoherent tunneling, the effect of the
tunneling Hamiltonian would be second order in t since the
expectation value of $S_x$ in Eq.(\ref{H_t}) is also proportional
to t. Therefore, for spontaneously coherent systems, the
interlayer tunneling conductance does not depend on t in the limit
of zero t while the charge conservation condition simply imposes
an upper bound on the value of coherent tunneling current, which
is proportional to $t$. In other words, our theory is consistently
constructed up to the first order of $t$, which is contrasted to
the second order behavior of the usual theory for incoherent
tunneling conductance.

Now, to proceed further, let us examine $H'_T$ in Eq.(\ref{H'_T})
in detail. Since the bilayer quantum Hall state is incompressible
at sufficiently small $d/l_B$, adding or removing electrons costs
a finite energy, $\Delta$, \cite{edge} which is equal to either
(i) the Coulomb self-energy of a quasiparticle, $\Delta_C$, for
the spontaneous coherence occurring when $t \ll e^2/\epsilon l_B$,
or (ii) the single-particle tunneling gap, $t$, for the induced
coherence occurring when $t \gg e^2/\epsilon l_B$. While it is
possible to investigate both regimes, the spontaneous coherence is
particularly interesting, as seen in later sections. It is,
however, sufficient at this stage to know that, no matter whether
$\Delta$ is $\Delta_C$ or $t$, $\Delta$ is independent of the
momentum index, $m$, in the lowest Landau level. So we just
replace $H_0+H_R+H_L-E_g$ by $\Delta$.

Note that we are able to make the above simplification because
$H'$ in Eq.(\ref{H'}) is chosen such that electrons are inserted
directly into Landau level eigenstates with momentum indices $m$.
Since each Landau-level eigenstate remains to be a well-defined
quasiparticle eigenstate in the presence of the Coulomb
interaction, one may take each $m$ state to be a very good
approximation to the exact eigenstates of $H_0+H_R+H_L$ with
eigenenergy $E_g +\Delta$ (Note that $\Delta$ is a self-energy
correction).

\begin{figure}
\includegraphics[width=1.4in]{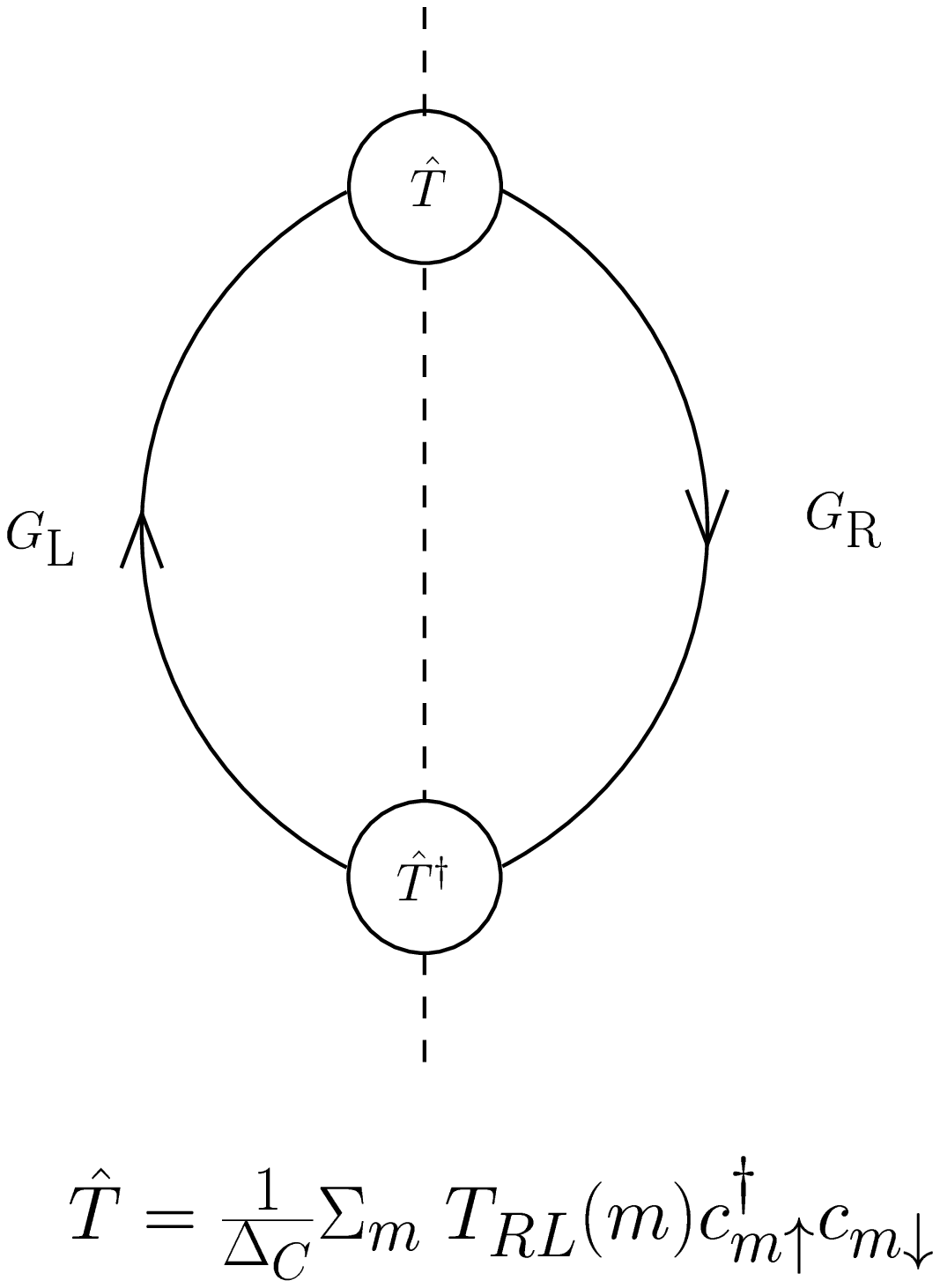}
\caption{Feynman diagram for coherent tunneling in bilayer quantum
Hall systems. The vertex operator $\hat{T}$ contains many-body
effects of the exciton condensate. $T_{RL}$ is the tunneling
amplitude and $\Delta_C$ is the Coulomb self-energy of a
quasiparticle. \label{feynman}}
\end{figure}

Now, we assume that the tunneling amplitudes $T_{R\uparrow}(k,m)$
and $T_{L\downarrow}(p,m')$ are more or less independent of
momenta $k$ and $p$, which is a common practice in tunneling
theories where tunneling occurs only within a narrow region of
energy near Fermi surface. Keeping only the terms in $H'_T$
relevant for transporting electrons from one lead to the other, we
arrive at the following many-body tunneling Hamiltonian:
\begin{equation}
H_T = \sum_{k,p} \left[
c^{\dagger}_{R}(k)c_L(p) \hat{T}^{\dagger}
+c^{\dagger}_{L}(p)c_R(k) \hat{T}
\right],
\label{H_T}
\end{equation}
where
\begin{equation}
\hat{T}=\frac{1}{\Delta}\sum_{m} T_{RL}(m) c^{\dagger}_{m\uparrow}
c_{m\downarrow}
\end{equation}
and $T_{RL}(m)=T_{R\uparrow}(k_F,m)T_{L\downarrow}(k_F,m)$. Based
on $H_T$, the tunneling current operator $\hat{J}$ is defined as
follows:
\begin{equation}
\hat{J} = e i \sum_{k,p} \left[
c^{\dagger}_{R}(k)c_L(p) \hat{T}^{\dagger}
-c^{\dagger}_{L}(p)c_R(k) \hat{T}
\right].
\label{J}
\end{equation}

Assuming that the tunneling amplitude between leads and the
bilayer system is small, one can compute the expectation value of
current operator via a conventional first-order S-matrix
expansion:
\begin{equation}
I(t)= -i \int^{t}_{-\infty} d t'
\langle [\hat{J}(t), H_T(t')] \rangle .
\label{current}
\end{equation}
The new aspect of our tunneling theory is the vertex operator,
$\hat{T}$, which contains all of many-body effects of the exciton
condensate. Eq.(\ref{current}) can be evaluated further using the
Feynman diagram depicted in Fig.\ref{feynman}:
\begin{eqnarray}
I &=& 2e |\langle\hat{T}\rangle|^2 \sum_{k,p}
\int^{\infty}_{-\infty} \frac{d \varepsilon}{2\pi}
A_R(k,\varepsilon)A_L(p,\varepsilon+eV)
\nonumber \\
&\times&[f(\varepsilon)-f(\varepsilon+eV)]
\nonumber \\
&=& 4\pi e^2 D_R D_L |\langle\hat{T}\rangle|^2 V
\label{Ohm}
\end{eqnarray}
where $A_R$ ($A_L$) is the spectral function of the right (left)
lead, $f(\varepsilon)$ is  the usual Fermi-Dirac distribution
function, and $D_R$ ($D_L$) is the density of states at the Fermi
surface of right (left) lead. Note that we have used
$|\langle\hat{T}\rangle|^2$ in Eq.(\ref{Ohm}) instead of
$\langle\hat{T}^\dagger \hat{T}\rangle$ since
$|\langle\hat{T}\rangle|^2$ is a much more acute measure of the
spontaneous interlayer order in finite-size system diagonalization
studies while the two quantities generate identical results in the
thermodynamic limit.

Eq.(\ref{Ohm}) indicates that there is no DC Josephson effect in a
conventional sense since the conductance $G$ ($\equiv dI/dV
\propto |\langle\hat{T}\rangle|^2$) is finite. It is important to
know, however, that the coherent tunneling current is zero if
$\langle\hat{T}\rangle = 0$. Remember that $\langle\hat{T}\rangle$
measures the phase coherence between states with various
interlayer number differences, $N_{\textrm{rel}}$  (Note that
$\hat{T} \propto c^{\dagger}_{m\uparrow} c_{m\downarrow}$). So,
unless the ground state is a coherent linear combination of states
with various $N_{\textrm{rel}}$, $\langle\hat{T}\rangle$ is zero,
and so is the tunneling current. As mentioned before, this is
precisely analogous to the phase coherence between different
number eigenstates in superconductivity, which is responsible for
the Josephson effect. In this sense, the interlayer tunneling
conductance is related to the Josephson effect. However, we
emphasize that the conductance is fundamentally finite even at
zero temperature. In the next section, we will actually evaluate
the interlayer tunneling conductance as a function of $d/l_B$ by
using numerical methods. In particular, we will be interested in
the normalized conductance since the absolute scale of the
conductance is sensitive to sample-specific details such as $D_R$,
$D_L$ and $T_{\textrm{RL}}$.

In concluding this section, we mention that the basic idea in our
derivation of the many-body tunneling Hamiltonian is rather
similar to that of the linear response theory; if a perturbation
is small enough, one can extract various dynamic properties of the
biased system from its ground state properties in equilibrium. In
our manuscript, we make a connection between the pseudospin order
parameter of the ground state and the tunneling conductance of the
steady state. This connection can be valid as long as the
interlayer current is not too large.

\section{Numerical Evaluation of the coherent tunneling current}
\label{numerics}

\begin{figure}
\includegraphics[width=2.7in,angle=0]{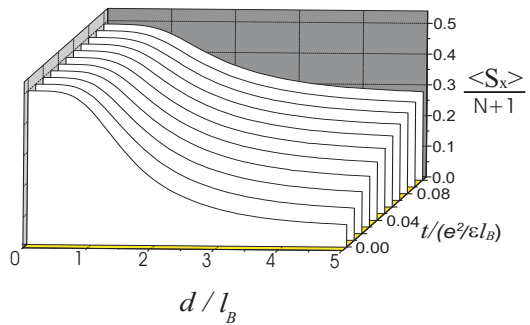}
\caption{Interlayer coherence quantified by the pseudospin
magnetization, $\langle S_x \rangle$, for the $N=11$ system.
$\langle S_x \rangle$ is computed via exact diagonalization as a
function of the interlayer distance, $d/l_B$, and the
single-particle tunneling parameter, $t/(e^2/\epsilon l_B)$. Note
that the shape of $\langle S_x \rangle$ is qualitatively similar
to the usual magnetization curve as a function of temperature and
external magnetic field, which correspond to interlayer distance,
$d$, and single-particle tunneling gap, $t$, respectively. Also,
note that $\langle S_x \rangle$ in the plot shows typical
finite-system smearing of a sharp phase transition. The
thermodynamic limit is estimated in Fig.\ref{Sxlimit} for
$t/(e^2/\epsilon l_B)=0$. \label{Sx}}
\end{figure}
To study the $d/l_B$ dependence of the tunneling conductance, one
has to compute $|\langle\hat{T}\rangle|^2$ in Eq.(\ref{Ohm}),
which can be further reduced as follows:
\begin{equation}
|\langle\hat{T}\rangle|^2 = \frac{\langle S_x \rangle^2}{\Delta^2}
\left| \frac{1}{N} \sum_m T_{\textrm{RL}}(m) \right|^2 ,
\label{T2}
\end{equation}
where $N$ is the total number of electrons. Here, we have
simplified the evaluation of the many-body tunneling matix
element, $\hat{T}$, by assuming that the interlayer phase
coherence is uniformly obtained throughout the system; in other
words, the interlayer coherence order parameter $\langle
c^{\dagger}_{m\uparrow} c_{m\downarrow} \rangle$ is independent of
m.

In this situation, the $m$-dependence of tunneling amplitudes
between leads and correspoding adjacent layers,
$T_{\textrm{RL}}(m)$, affects only the overall magnitude of the
tunneling conductance, leaving its $d/l_B$ dependence unchanged.
Note that effectively what we are assuming is that the lead does
not alter important bilayer properties such as the interlayer
distance and the single-particle tunneling amplitude. Now, since
$\langle c^{\dagger}_{m\uparrow} c_{m\downarrow} \rangle$ is
independent of m, $\langle c^{\dagger}_{m\uparrow} c_{m\downarrow}
\rangle = \langle S_x \rangle /N $. $S_x$ [$= \sum_m
(c^{\dagger}_{m\uparrow} c_{m\downarrow}
+c^{\dagger}_{m\downarrow} c_{m\uparrow})/2$] can be regarded as
an order parameter of the exciton condensate. It is natural to
assume that $\sum_m T_{\textrm{RL}}(m)/N$ does not depend on
$d/l_B$ since $T_{\textrm{RL}}$ is only dependent on the hopping
amplitude between lowest-Landau-level states in a given layer and
plane-wave states in the external lead connected to the layer. The
$d/l_B$ dependence of the conductance is, therefore, solely
determined by $\langle S_x \rangle^2/\Delta^2$.

Figure \ref{Sx} shows the three-dimensional plot of $\langle S_x
\rangle$ as a function of the interlayer distance, $d/l_B$, and
the single-particle tunneling gap, $t/(e^2/\epsilon l_B)$.
$\langle S_x \rangle$ in the plot was computed via exact
diagonalization for the $N=11$ system. Note that, when computing
$\langle S_x \rangle$ in finite systems, it is very important to
take into account fundamental fluctuations in $N_{\textrm{rel}}$;
the true ground state is a coherent, linear combination of states
with various $N_{\textrm{rel}}$. Technical details can be found in
the literature \cite{Park,Park2,NJP}. Also, note that $\langle S_x
\rangle$ in the plot shows typical finite-system smearing of a
sharp phase transition. It is interesting that the shape of
$\langle S_x \rangle$ is qualitatively similar to the typical
magnetization curve as a function of temperature and external
magnetic field, which respectively correspond to $d/l_B$ and
$t/(e^2 /\epsilon l_B)$.
\begin{figure}
\includegraphics[width=2.2in,angle=0]{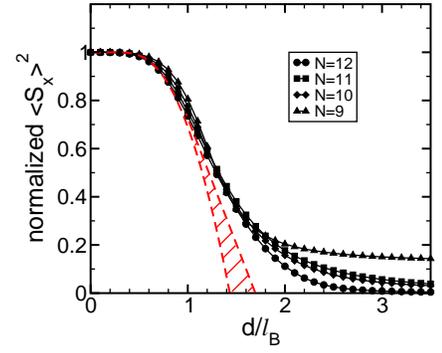}
\caption{Normalized expectation value of the condensate order
parameter ${\langle S_x \rangle}^2$ in the limit of $t/(e^2
\epsilon l_B) \rightarrow 0$. In order to incorporate
finite-system corrections, $\langle S_x \rangle$ is normalized in
such a way that it is divided by $(N+1)/2$ for $N$ odd and by
$\sqrt{N(N+2)}/2$ for $N$ even \cite{Park}. $N$ is the total
number of electrons in finite-system exact diagonalization
studies. The shaded region indicates the thermodynamic-limit
estimate for ${\langle S_x \rangle}^2$.
\label{Sxlimit}}
\end{figure}

Figures \ref{Sx} and \ref{Sxlimit} show that, for $d < d_C \simeq
1.4 - 1.7 l_B$, $\langle S_x \rangle$ does not vanish even in the
limit of $t/(e^2/\epsilon l_B) \rightarrow 0$, which is, by
definition, the evidence for the spontaneous order. Since $\langle
S_x \rangle$ becomes zero for $d > d_C$ in the thermodynamic
limit, it is natural that the tunneling conductance becomes zero
as well, as dictated by Eqs.(\ref{Ohm}) and (\ref{T2}). On the
other hand, if induced by the single-particle tunneling alone (or,
if $t \gg e^2/\epsilon l_B$), the coherence does not completely
vanish at a finite $d/l_B$, nor does the corresponding coherent
tunneling conductance peak. This is not consistent with experiment
where the conductance peak disappears at a finite critical
interlayer distance.

So, from now on, we focus on the limit of very small
single-particle tunneling gap, $t/(e^2/\epsilon l_B) \rightarrow
0$, where the coherence is spontaneous. Fig.\ref{Sxlimit} plots
$\langle S_x \rangle^2$ as a function of $d/l_B$ for various
particle numbers when $t/(e^2/\epsilon l_B) \rightarrow 0$.
While the accurate estimate of the thermodynamic limit of $\langle
S_x \rangle^2$ is difficult due to numerical uncertainties, it is
reasonable to assume that the true thermodynamic limit lies within
the shaded region in Fig.\ref{Sxlimit}. As typical in
finite-system calculations, inflection points are taken to be the
finite-system signature for the true critical point in the
thermodynamic limit.

\begin{figure}
\includegraphics[width=2.2in,angle=0]{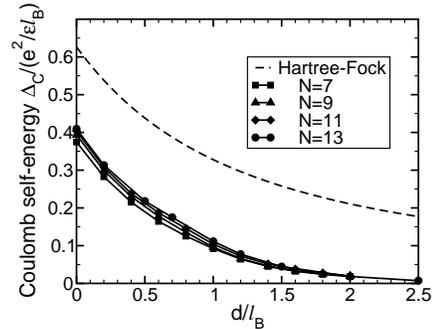}
\caption{Energy cost of adding (removing) an electron into (from)
the ground state of the bilayer quantum Hall system. Note that
this energy cost is essentially the self-energy of a quasiparticle
due to the Coulomb interaction. The Coulomb self-energy is
computed via exact diagonalization (for various finite systems) as
a function of interlayer distance, $d/l_B$. For comparison, also
plotted is the self-energy estimate obtained in the Hartree-Fock
approximation. \label{DeltaC}}
\end{figure}
In Fig.\ref{DeltaC}, we compute the Coulomb energy cost of
creating a quasiparticle, $\Delta_C$, which is the only energy
cost when $t/(e^2/\epsilon l_B) \rightarrow 0$. That is, $\Delta =
\Delta_C$ in Eq.(\ref{T2}). $\Delta_C$ is computed via exact
diagonalization in finite systems. It is interesting to note that,
for general values of $d/l_B$, the exact $\Delta_C$ is much lower
than the estimate obtained in the Hartree-Fock approximation
\cite{Fertig,MPB}.

\begin{figure}
\includegraphics[width=2.4in,angle=0]{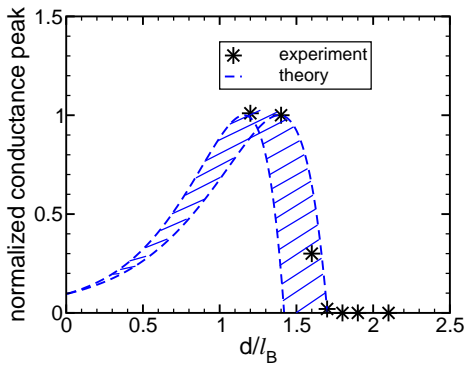}
\caption{Normalized interlayer tunneling conductance peak (the
shaded region in the plot) as a function of $d/l_B$ in comparison
with experimental data from Spielman {\it et al.} \cite{Spielman}.
We define the normalized conductance as the conductance divided by
its maximum value as a function of $d/l_B$. The shaded region is
obtained from the thermodynamic estimate of $\langle S_x
\rangle^2$ in Fig.\ref{Sxlimit}. \label{conductance}}
\end{figure}

Finally, in Fig.\ref{conductance}, our theoretical estimate for
the normalized interlayer tunneling conductance near zero bias,
{\it i.e.} $\langle S_x \rangle^2/\Delta^2_C$, is compared with
experimental data of Spielman {\it et al.} \cite{Spielman}. We
define the normalized conductance as the conductance divided by
its maximum value as a function of $d/l_B$. The shaded region in
the plot is the thermodynamic estimate.
To the best of our knowledge, the direct comparison with
experimental data for the conductance peak is made here for the
first time. It is interesting that our theory predicts that the
conductance peak decreases as $d/l_B$ decreases below roughly 1.2.
Remember that the decrease in the conductance peak at small
$d/l_B$ is due to the increase in $\Delta_C$ while the pseudospin
magnetization is saturated.

We would like to emphasize that our theoretical estimate of the
conductance peak has only one fitting parameter: the overall
prefactor dependent on sample specifics. Considering that the only
fitting parameter is an overall scale factor, we argue that the
agreement with experiment is not too bad. Specifically, our theory
provides a reasonably accurate estimate for the critical distance
at which the conductance peak disappears. While the critical
distance was estimated previously, its determination has been
based on indirect evidence: it was determined either (i) by
studying the collapse of the overlap between the Halperin's
(1,1,1) state and the exact ground state \cite{SongHe,Yoshioka},
or (ii) by studying the collapse of the low-energy excitation in a
time-dependent Hartree-Fock approximation \cite{MPB} or (iii) by
approaching the critical point from the incoherent side and
studying the effect of the single-particle tunneling
\cite{Schliemann}. Our theory, on the other hand, is a direct
study of the tunneling conductance itself, as measured
experimentally, rather than a consideration of an energy gap
collapse or similar purely theoretical constructs associated with
the quantum phase diagram.¡±

Aside from the direct estimation of the critical distance, our
theory predicts a rather steep rise of the peak height as $d/l_B$
is reduced from the critical value, as seen in Fig.
\ref{conductance}. This steep rise cannot be explained by the
collapse of the pseudospin magnetization, $\langle S \rangle$,
alone; it requires a detailed consideration of the coherent
tunneling process.

In the preceding section, we have shown that there is a critical
interlayer current allowed without breaking the phase coherence.
Consequently, for a sufficiently large bias voltage, the coherent
interlayer current should be cut off. We have argued that this is
the reason why the interlayer conductance has a shape of the sharp
peak near zero bias. Now, we would like to discuss what determines
the width of the conductance peak. By equating
Eqs.(\ref{Jcritical}) and (\ref{Ohm}), one can show that, while
the proportionality constant strongly depends on sample-specific
details, the width of conductance peak is proportional to $t$. Due
to the strong dependence on sample details, the accurate
estimation of the critical current is beyond the scope of this
paper. It is, however, encouraging to find that typical width of
the conductance peak ($\sim 10 \mu eV$) is in the similar order as
the single-particle tunneling gap \cite{Spielman,Spielman2}. Our
finding that the width of the conductance peak is proportional to
the tunneling amplitude $t$ shows that the existence of a finite
symmetry breaking term, i.e. a nonzero $t$, is essential for the
experimental observations in Spielman {\it et al.} \cite{Spielman}

\section{Proposal for the Josephson effect in conterflowing current measurements}
\label{realJosephson}

Until now, we have studied interlayer tunneling in a single
bilayer system, which, we showed, is not an exact analog of the
Josephson effect. In this section we propose a much more direct
analog of the real Josephson effect. Since the coherent
particle-number fluctuation in superconductivity holds a parallel
with the coherent fluctuation of interlayer number differences in
bilayer quantum Hall systems, it is natural to expect that the
corresponding Josephson effect for the bilayer quantum Hall
systems must manifest itself in current measurements associated
with interlayer number difference. Such measurements are
counterflowing current measurements\cite{Kellogg,Shayegan}.

\begin{figure}
\includegraphics[width=3in,angle=0]{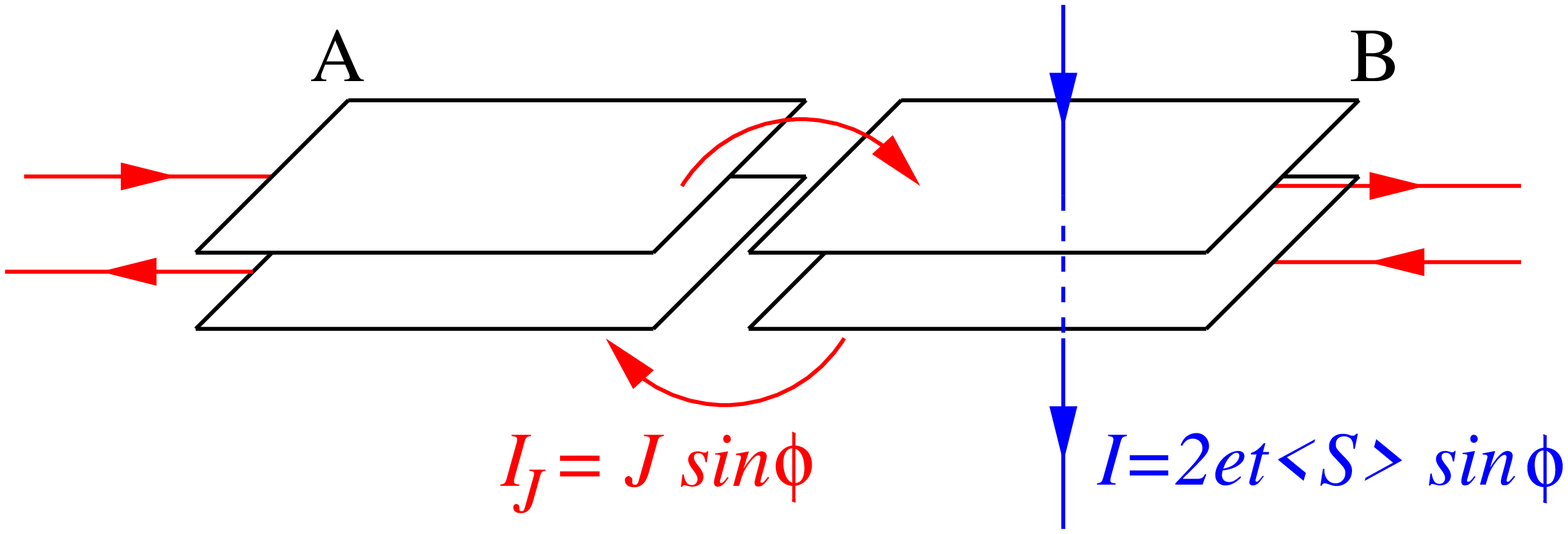}
\caption{Schematic diagram for the proposed Josephson effect in
counterflowing current measurements in a coupled set of two
bilayer quantum Hall systems. \label{Josephson}}
\end{figure}

To be specific, let us consider a weakly coupled set of two
bilayer quantum Hall systems, say A and B (four layers
altogether), separated by a lateral tunneling barrier. We further
consider the situation where there is no cross tunneling between
the top layer of the system A and the bottom layer of the system
B. Tunneling occurs either (i) between the top and the bottom
layer of the same system or (ii) between the same layers of the
system A and B. A schematic diagram is shown in
Fig.\ref{Josephson}.

The configuration similar to the one depicted in
Fig.\ref{Josephson} has been, in fact, proposed previously
\cite{WenZee_Josephson}. The important twist in this paper is the
application of the interlayer current flowing vertically through
the top and the bottom layer of one of the bilayer systems.  The
purpose is to induce a non-zero interlayer phase difference in one
of the bilayer systems while the other system has none. Note that,
in bilayer quantum Hall systems, the phase of the condensate is
controlled via vertical interlayer current flow which affects the
internal degree of freedom of the condensate. It is interesting
that there is no similar way of controlling the phase in usual
superconductors.

Specifically, when the applied interlayer current is $I$, a
non-zero interlayer phase is induced so that
\begin{equation}
\sin{\phi}= I/2et\langle S \rangle,
\end{equation}
where  $\langle S \rangle$ is defined in Eq.(\ref{Jinter}). Our
prediction then is that, due to the analogy with the Josephson
effect in superconductivity, there will be two counterflowing
currents with one flowing between the top layers of the system A
and B, and with the other flowing between the bottom layers; the
Josepson effect for the neutral, total current in our system is
given by
\begin{equation}
I_J = J \sin{\phi} = \frac{J}{2et\langle S \rangle} I,
\end{equation}
where the proportionality constant, $J$, is dependent on tunneling
parameters between the system A and B. Note that, when $I_J$ is
the current flowing between the top layers, the current of $-I_J$
flows between the bottom layers. The net current is zero, but it
may be possible to measure these two counterflowing currents
individually.

Let us mention what conditions should be satisfied for our
prediction to be realized. In the above, we have already mentioned
that the cross tunneling between different layers of the system A
and B is assumed to be negligible. This is due to the fact that
the cross tunneling cannot carry the supercurrent. A maybe more
important condition, however, is that the applied current, $I$,
should not exceed the critical current since, otherwise, the
coherence will be lost. The experiment described in the above,
therefore, can be performed only in a very narrow range of the
applied current, which, in turn, corresponds to a very narrow
bias-voltage range.

\section{Conclusion}
\label{conclusion}

It has been shown in the preceding sections that the interlayer
tunneling conductance peak near zero bias is strongly enhanced,
but fundamentally finite even at zero temperature. The reason why
the interlayer conductance peak is not intrinsically infinite can
be attributed to the fact that the experimental setup measuring
the interlayer tunneling conductance is not a setup for the true
Josephson effect in the bilayer exciton condensate. To
substantiate this idea, we have computed the height of the
zero-bias interlayer conductance peak as a function of interlayer
distance, which is, then, compared with experiment. Considering
that the only fitting parameter is an overall scale factor which
strongly depends on sample specifics, the agreement with
experiment is reasonable. Specifically, our theory provides a
reasonably accurate estimate for the critical distance in which
the conductance peak disappears. Our theory is contrasted from
previous theories in that it determines the critical distance by
explicitly computing the interlayer tunneling conductance as a
function of interlayer distance. Thus, our theory is a transport
theory for the tunneling conductance, which is precisely what is
measured experimentally.

While one of the main goals of this paper is to develop a concrete
theory of the coherent tunneling conductance, it is another goal
of this paper to make a sharp distinction between the spontaneous
coherence due to the Coulomb interaction and the induced coherence
due to the single-particle tunneling. In this respect, the
detailed computation of the conductance peak height as a function
of $d/l_B$ is also very important because it offers such sharp
distinction; if the coherence is due to the Coulomb interaction,
the conductance peak will completely disappear at a finite
$d/l_B$, as discussed in Sec.\ref{coherent}.

We have also discussed the similarities and the differences
between the bilayer quantum Hall interlayer coherence phenomena
and the related collective coherent physics manifested in magnetic
phenomena and in superconducting Josephson effect. In particular,
we have proposed an experimental setup for the Josephson effect in
couterflowing current measurements in a coupled set of two bilayer
quantum Hall systems, which is a precise analogy with the real
Josephson effect in superconductivity.

\section{Acknowledgement}

The authors are especially grateful to J. P. Eisenstein for
careful reading of the manuscript and valuable suggestions. The
authors are also indebted to J. K. Jain, Y. N. Joglekar, Y.-B.
Kim, Yong-Hoon Kim, and V. W. Scarola for their insightful
comments. This work was supported by LPS-NSA and ARO-ARDA.



\begin{thebibliography}{}
\bibitem{edge} Bilayer quantum Hall systems have
an edge excitation which requires an infinitesimally small energy
to be created. However, the edge excitation does not have any
contribution to coherent interlayer tunneling since it is not a
part of coherent exciton condensate and its contribution to
interlayer tunneling is only through single-particle tunneling
which is of the order of extremely small single-particle tunneling
gap $t$.
\end{thebibliography}
\end{document}